\documentclass[twocolumn]{revtex4}
\usepackage{amsmath,amssymb}
\usepackage{graphicx}
\usepackage{dcolumn}
\usepackage[dvips]{epsfig,color}
\usepackage{bm}
\begin{document}
\title{Spectroscopy of higher bottomonia\footnote{Talk given at the conference "Dark Matter, Hadron Physics and Fusion Physics", Messina (Italy), September 24-26, 2014}}

\author{J. Ferretti}
\affiliation{Dipartimento di Fisica and INFN, Universit\`a di Roma "Sapienza", Piazzale A. Moro 5, I-00185 Roma, Italy}

\begin{abstract}
In this contribution, we discuss our recent unquenched quark model results for the spectrum of $b \bar b$ mesons with self energy corrections, due to the coupling to the meson-meson continuum. 
Our unquenched quark model predictions for the masses of the recently discovered $\chi_b(3P)$ states are compared to those of a re-fit of Godfrey and Isgur's relativized quark model to the most recent experimental data.  The possible importance of continuum effects in the $\chi_b(3P)$ states is discussed. Finally, we show our quark model results for the radiative decays of the $\chi_b(3P)$ system and the open-bottom decays of $b \bar b$ mesons.
\end{abstract}

\maketitle

\section{Introduction. An unquenched quark model for mesons}
In the UQM for hadrons \cite{bottomonium,Ferretti:2013vua,Bijker:2009up,Santopinto:2010zza,Bijker:2012zza,charmonium,charmonium02}, the effects of $q \bar q$ sea pairs are introduced explicitly into the quark model (QM) via a QCD-inspired $^{3}P_0$ pair-creation mechanism \cite{3P0}. 
The $q \bar q$ pairs have vacuum quantum numbers and their effect on hadron observables can be added as a perturbation.
This approach is a generalization of the unitarized quark model by T\"ornqvist and Zenczykowski \cite{Tornqvist} and was also motivated by the work of Geiger and Isgur \cite{Geiger-Isgur}.

In the UQM for mesons \cite{bottomonium,Ferretti:2013vua,charmonium,charmonium02}, the meson wave function,
\begin{eqnarray} 
	\label{eqn:Psi-A}
	\mid \psi_A \rangle &=& {\cal N} \left[ \mid A \rangle 
	+ \sum_{BC \ell J} \int d \vec{q} \, \mid BC \vec{q} \, \ell J \rangle \right.
	\nonumber\\
	&& \hspace{2cm} \left.  \frac{ \langle BC \vec{q} \, \ell J \mid T^{\dagger} \mid A \rangle } 
	{E_a - E_b - E_c} \right] ~, 
\end{eqnarray}
is written as the superposition of a zeroth order quark-antiquark configuration and a sum over the possible higher Fock components, due to the creation of $^{3}P_0$ $q \bar q$ pairs. 
Above threshold, this coupling to the continuum leads to open-flavor strong decays; below threshold, it leads to virtual $q \bar q - q \bar q$ components in the meson wave function and a shift of the physical meson mass with respect to its bare mass, namely a self-energy correction. 
Something similar was also done in the baryon sector \cite{Bijker:2009up,Santopinto:2010zza,Bijker:2012zza}, with the intention to extend the range of applicability of the baryon QM in its various forms \cite{Isgur:1979be,Capstick:1986bm,Bijker:1994yr,Glozman-Riska,Loring:2001kx,Giannini:2001kb,Bijker:2004gr,Ferretti:2011,Galata:2012xt,Aznauryan:2012ba}.

In Ref. \cite{bottomonium}, the UQM formalism was used to calculate the mass shifts of $1S$, $2S$ and $1P$ bottomonium states due to the coupling to the meson-meson continuum. 
This was just a preliminary study, in which the bare energies were taken as free parameters and their sum with the self-energies fitted to the physical masses of the mesons of interest. 
There, we also calculated the probability to find the wave function of $1S$, $2S$ and $1P$ bottomonium states in the continuum component. The results vary between $2\%$ and $6\%$ and their smallness is due to the distance between these low-lying states and the first open-bottom decay thresholds.  

In Ref. \cite{charmonium}, the UQM was used to calculate the charmonium spectrum ($1S$, $2S$, $1P$, $2P$ and $1D$ states) with self-energy corrections. Unlike Ref. \cite{bottomonium}, the bare energies were calculated explicitly within a potential model \cite{Godfrey:1985xj}, thus extending the predictive power of the UQM. 
Our results were used to discuss the nature of the $X(3872)$ \cite{Choi:2003ue}, a meson characterized by $J^{PC} = 1^{++}$ quantum numbers \cite{Aaij:2013zoa}, a very narrow width (less than 1.2 MeV \cite{Nakamura:2010zzi}), and a mass $50-100$ MeV lower than QM predictions. 
According to our study, the $X(3872)$ can be interpreted as a $c \bar c$ core plus higher Fock components due to the coupling to the meson-meson continuum. 
The problem of the mass difference between QM predictions \cite{Godfrey:1985xj,Eichten:1978tg,Barnes:2005pb} and the experimental data for the $X(3872)$ was solved thanks to a downward mass shift, due to self-energy corrections. See also Ref. \cite{charmonium02}.

In Ref. \cite{Ferretti:2013vua}, this calculation was extended to bottomonia and used to provide predictions for $1S$, $2S$, $3S$, $1P$, $2P$, $3P$ and $1D$ resonances; the open-bottom decays of $b \bar b$ states were calculated too within the $^3P_0$ pair-creation model \cite{3P0}.
Analogously to what happens for the $X(3872)$ in the $c \bar c$ sector, threshold effects may play an important role in the $\chi_b(3P)$ system, discovered by ATLAS in 2012 \cite{Aad:2011ih,Abazov:2012gh}, because $\chi_b(3P)$ mesons are very close to the first open-bottom decay thresholds.
Thus, to distinguish between QM and UQM interpretations for these mesons, our UQM results for the mass barycenter of the $\chi_b(3P)$ system were compared to those of a re-fit of Godfrey and Isgur's model mass formula to the most recent experimental data.  
In Ref. \cite{charmonium02}, the radiative transitions of $\chi_b(3P)$ mesons were calculated too.

In this contribution, we briefly analyze the results of Refs. \cite{Ferretti:2013vua,charmonium02} for the bottomonium spectrum, the open-bottom decays and the radiative transitions of $\chi_b(3P)$ mesons. These results are then used to discuss the possible importance of continuum effects in $\chi_b(3P)$ mesons.

%%%%%%%%%%%%%%%%%%%%%%%%%%%%%%%%%%%%%%%%
\begin{figure}[htbp]
\begin{center}
\includegraphics[width=8cm]{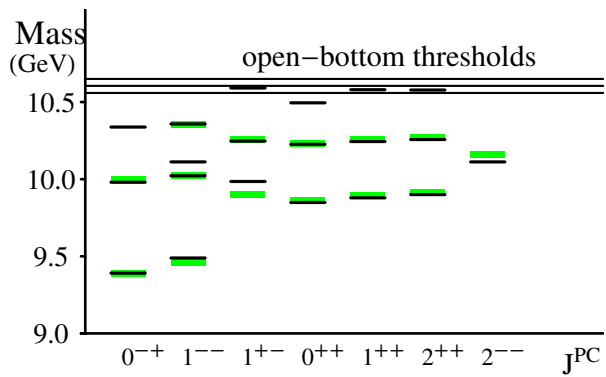}
\end{center}
\caption{Spectrum of $b \bar b$ mesons in the UQM. Picture from Ref. \cite{Ferretti:2013vua}; APS copyright. The calculated masses of $b \bar b$ states are shown as black lines, the experimental data \cite{Nakamura:2010zzi} as green boxes.} 
\label{fig:bottomonium-spectrum-self}
\end{figure} 
%%%%%%%%%%%%%%%%%%%%%%%%%%%%%%%%%%%%%%%%

\section{Spectroscopy of higher bottomonia}
\subsection{Bottomonium spectrum in the UQM}
Here, we discuss our UQM results for the spectrum of $b \bar b$ mesons with self energy corrections of Refs. \cite{bottomonium,Ferretti:2013vua}.
In our calculation, we fitted 
\begin{equation}
	\label{eqn:self-trascendental}
	M_a = E_a + \Sigma(E_a)  
\end{equation}
to the experimental data \cite{Nakamura:2010zzi}.
The physical mass of a meson, $M_a$, results from the sum of a bare energy term, $E_a$, computed within Godfrey and Isgur's relativized QM \cite{Godfrey:1985xj}, and a self energy correction, 
\begin{equation}
	\label{eqn:self-a}
	\Sigma(E_a) = \sum_{BC} \int_0^{\infty} q^2 dq \mbox{ } \frac{\left| \left\langle BC \vec q  \, \ell J \right| T^\dag \left| A \right\rangle \right|^2}{E_a - E_{bc}}  \mbox{ },
\end{equation}
evaluated in the unquenched quark model formalism \cite{bottomonium,Ferretti:2013vua,Bijker:2009up,charmonium,charmonium02}. See Fig. \ref{fig:bottomonium-spectrum-self}. 
In Ref. \cite{Ferretti:2013vua}, we also gave results obtained by re-fitting the relativized QM mass formula \cite{Godfrey:1985xj} to the most recent experimental data \cite{Nakamura:2010zzi}, without including continuum effects.
The re-fit was necessary to compute the open-bottom strong decays, which required more precise values for the masses of the higher-lying decaying mesons, like $4S$ states.

It is interesting to notice that (see Refs. \cite{Ferretti:2013vua,charmonium02}): 1) below 10.5 GeV, our UQM and relativized QM results are very similar; 2) above 10.5 GeV, an evident discrepancy emerges in the case of $\chi_b(3P)$ resonances \cite{Aad:2011ih,Abazov:2012gh}, due to the vicinity of these states to the first open-bottom decay thresholds. 
This means that, far from thresholds, continuum effects can be (at least partially) re-absorbed into a redefinition of the model parameters, as also demonstrated in Ref. \cite{Geiger:1989yc}; close to thresholds, this is not possible anymore and continuum effects can play an important role. 
For example, this also happens in the case of the $X(3872)$ \cite{Pennington:2007xr,Li:2009ad,Danilkin:2010cc,charmonium,charmonium02} and $D_{sJ}^*(2317)$ \cite{Hwang:2004cd} mesons.

%%%%%%%%%%%%%%%%%%%%%%%%%%%%%%%%%%%%%%%
\begin{table}[htbp]  
\begin{center}
\begin{tabular}{lll}
\hline
$M^{\mbox{th}}_{\chi_b(1P)}$ &  $\Delta M_{21}(1P)$ & $\Delta M_{10}(1P)$    \\
9876                         &  21                  & 30  \\
\hline
$M^{\mbox{th}}_{\chi_b(2P)}$ &  $\Delta M_{21}(2P)$ & $\Delta M_{10}(2P)$    \\
10242                        &  13                  & 18  \\
\hline
$M^{\mbox{th}}_{\chi_b(3P)}$ &  $\Delta M_{21}(3P)$ & $\Delta M_{10}(3P)$    \\
10551                        &  -2                  & 85 \\ 
\hline
\end{tabular}
\end{center}
\caption{Mass barycenters of $\chi_b(1P)$, $\chi_b(2P)$ and $\chi_b(3P)$ systems (column 1) and mass splittings between the members of the $\chi_b(1P)$, $\chi_b(2P)$ and $\chi_b(3P)$ multiplets (column 2 and 3) in the UQM. See Ref. \cite{Ferretti:2013vua}. The results are expressed in MeV. The notation $\Delta M_{21}(1P)$ stands for the mass difference between the $\chi_{b2}(1P)$ and $\chi_{b1}(1P)$ resonances, $\Delta M_{10}(1P)$ for the mass difference between the $\chi_{b1}(1P)$ and $\chi_{b0}(1P)$ resonances, and so on.}
\label{tab:mass-bary}
\end{table}

\begin{table}[htbp]  
\begin{center}
\begin{tabular}{lll}
\hline
$M_{\chi_b(1P)}$ & $\Delta M_{21}(1P)$ & $\Delta M_{10}(1P)$    \\
9894             & 21                  & 30     \\ 
\hline
$M_{\chi_b(2P)}$ & $\Delta M_{21}(2P)$ & $\Delta M_{10}(2P)$    \\
10241            & 21                  & 14      \\ 
\hline
$M_{\chi_b(3P)}$ & $\Delta M_{21}(1P)$ & $\Delta M_{10}(1P)$    \\
10510            & 17                  & 13      \\ 
\hline
\end{tabular}
\end{center}
\caption{Mass barycenters of $\chi_b(1P)$, $\chi_b(2P)$ and $\chi_b(3P)$ systems (column 1) and mass splittings between the members of the $\chi_b(1P)$, $\chi_b(2P)$ and $\chi_b(3P)$ multiplets (column 2 and 3), from our re-fit of Godfrey and Isgur's mass formula of Ref. \cite{Ferretti:2013vua}. The results are expressed in MeV.}
\label{tab:mass-bary-GI}
\end{table}

\begin{table}
\centering
\begin{tabular}{lll}
\hline
$M_{\chi_c(1P)}$ & $\Delta M_{21}(1P)$ & $\Delta M_{10}(1P)$    \\
3494             & 46                  & 96     \\ 
\hline
$M_{\chi_c(2P)}$ & $\Delta M_{21}(2P)$ & $\Delta M_{10}(2P)$    \\
3906             & 56                  & -47     \\ 
\hline
\end{tabular}
\caption{Experimental data \cite{Nakamura:2010zzi} for the mass barycenters of $\chi_c(1P)$ and $\chi_c(2P)$ systems (column 1) and mass splittings between the members of the $\chi_c(1P)$ and $\chi_c(2P)$ multiplets (column 2 and 3). The results are expressed in MeV.}
\label{tab:mass-bary-ChiC}
\end{table}

\begin{table*}
\centering
\begin{tabular}{lllllllll} 
\hline
Meson                            &  Mass [MeV] &  $J^{PC}$    & $B \bar B$ & $B \bar B^*$ / $\bar B B^*$ & $B^* \bar B^*$ & $B_s \bar B_s$ & $B_s \bar B_s^*$ / $\bar B_s B_s^*$ & $B_s^* \bar B_s^*$ \\ 
\hline
$\Upsilon(10580)$ or $\Upsilon(4^3S_1)$      & 10.595                            & $1^{--}$   & 20 & -- & -- &  --   &  --  &  --  \\
                                    & $10579.4\pm1.2^\dag$ &                  &       &     &     &         &        &        \\
$\chi_{b2}(2^3F_2)$     & $10585$ & $2^{++}$ & 34  & -- & -- &  --  &  --  &  --  \\
$\Upsilon(3^3D_1)$     & $10661$ & $1^{--}$   & 23 &  4  & 15 &  --  &  --  &  --  \\
$\Upsilon_2(3^3D_2)$ & $10667$ & $2^{--}$ &   -- & 37  & 30 &  --  &  --  &  --  \\
$\Upsilon_2(3^1D_2)$ & $10668$ & $2^{-+}$ &  --  & 55  & 57 &  --  &  --  &  --  \\  
$\Upsilon_3(3^3D_3)$ & $10673$ & $3^{--}$ &  15  & 56  &113 &  -- &  --  &  --  \\ 
$\chi_{b0}(4^3P_0)$     & $10726$ & $0^{++}$ & 26  & -- & 24 &  --  &  --  & --  \\
$\Upsilon_3(2^3G_3)$ & $10727$ & $3^{--}$ &   3   &  43  & 39 &  -- &  --  &  -- \\
$\chi_{b1}(4^3P_1)$     & $10740$ & $1^{++}$ & --  & 20 &  1  &  --  &  --  &  --  \\
$h_b(4^1P_1)$             & $10744$ & $1^{+-}$ &  --  & 33 &  5  &  --  &  --  &  --  \\
$\chi_{b2}(4^3P_2)$     & $10751$ & $2^{++}$ & 10  & 28 &  5  &   1  &  --  &  --  \\
$\chi_{b2}(3^3F_2)$     & $10800$ & $2^{++}$ &  5   & 26 & 53 &   2  &   2   &  --  \\
$\Upsilon_3(3^1F_3)$  & $10803$ & $3^{+-}$ &  --  & 28   & 46  &  --  &  3 &  --  \\
$\Upsilon(10860)$ or $\Upsilon(5^3S_1)$      & $10876\pm11^\dag$ & $1^{--}$  &  1  & 21 & 45 &   0   &   3   &   1  \\
$\Upsilon_2(4^3D_2)$ & $10876$ & $2^{--}$ &  --  & 28  & 36 &  --  &   4   &   4   \\
$\Upsilon_2(4^1D_2)$ & $10877$ & $2^{-+}$ &  --  & 22  & 37 &  --  &   4   &   3   \\
$\Upsilon_3(4^3D_3)$ & $10881$ & $3^{--}$ &   1   &  4    & 49 &   0  &   1  &   2  \\
$\Upsilon_3(3^3G_3)$ & $10926$ & $3^{--}$ &   7   &  0    & 13 &   2   &   0  &  5   \\
$\Upsilon(11020)$ or $\Upsilon(6^3S_1)$      & $11019\pm8^\dag$ & $1^{--}$  &  0  &  8  & 26 &   0   &   0   &   2  \\   
\hline
\end{tabular} 
\caption{Open-bottom decay widths (in MeV) for higher bottomonium states, as from Ref. \cite{Ferretti:2013vua}; APS copyright. Column 2 shows the values of the masses of the decaying $b \bar b$ states: when available, we used the experimental values from PDG \cite{Nakamura:2010zzi} ($\dag$), otherwise the theoretical predictions of the relativized QM \cite{Godfrey:1985xj}, whose mass formula we re-fitted to the most recent experimental data \cite{Ferretti:2013vua}. Columns 3-8 show the decay width contributions from various $BC$ channels, like $B \bar B$, $B \bar B^*$ and so on. The values of the $^3P_0$ model parameters were fitted to experimental data for the strong decays of $b \bar b$ resonances \cite{Ferretti:2013vua}. The symbol -- in the table means that a certain decay is forbidden by selection rules or that the decay cannot take place because it is below the threshold.}
\label{tab:strong-decays}  
\end{table*}
%%%%%%%%%%%%%%%%%%%%%%%%%%%%%%%%%%%%%%%

\subsection{Mass barycenter of the $\chi_b(3P)$ system}
Our UQM results were also used to evaluate the mass barycenter of the $\chi_b(3P)$ system. See Table \ref{tab:mass-bary}. Another estimation was obtained by re-fitting the relativized QM mass formula \cite{Godfrey:1985xj} to the most recent experimental data \cite{Nakamura:2010zzi,Mizuk:2012pb}. 
See Table \ref{tab:mass-bary-GI}. 
The two sets of theoretical results can be then compared to the experimental data: $M_{\chi_b(3P)} = 10.534 \pm 0.009$ GeV \cite{Nakamura:2010zzi}, $M_{\chi_b(3P)} = 10.530 \pm 0.005 (\mbox{stat.}) \pm 0.009 (\mbox{syst.})$ GeV \cite{Aad:2011ih}, and $M_{\chi_b(3P)} = 10.551 \pm 0.014 (\mbox{stat.}) \pm 0.017 (\mbox{syst.})$ GeV \cite{Abazov:2012gh}.
LHCb has also recently provided a measure for the mass of the $\chi_{b1}(3P)$ resonance \cite{Aaij:2014hla}: $M_{\chi_{b1}(3P)} = 10515.7^{+2.2}_{-3.9}(\mbox{stat.})^{+1.5}_{-2.1}(\mbox{syst.})$ MeV.
See also Ref. \cite{Dib:2012vw}, where the authors provided predictions for the splittings between the members of the $\chi_b(3P)$ multiplet using several potential models.

It is worthwhile noting that: 1) our UQM and relativized QM predictions of Tables \ref{tab:mass-bary} and \ref{tab:mass-bary-GI} are similar in the case of the mass barycenter, which is the average between the three meson masses of the multiplet, but differ for the splittings between the members of the multiplet; 2) the mass splitting scheme, with $M_{\chi_{b1}(1P,2P)} - M_{\chi_{b0}(1P,2P)} \approx 15-30$ MeV and $M_{\chi_{b2}(1P,2P)} - M_{\chi_{b1}(1P,2P)} \approx 15-20$ MeV, in the UQM is broken for $\chi_b(3P)$ mesons. Something similar also happens for $\chi_c(2P)$ mesons. See Table \ref{tab:mass-bary-ChiC}. Therefore, we think that new and more precise experimental data will be necessary to confirm or discard our hypothesis of $\chi_b(3P)$ mesons as made up of a $b \bar b$ core plus non-negligible continuum components, due to threshold effects. 

\section{Open bottom strong decays in the $^3P_0$ pair-creation model}
\label{Strong decay widths} 
Here, we discuss our results for the open-bottom strong decays of Ref. \cite{Ferretti:2013vua}.
The transitions we considered are of the type
\begin{equation}
	\begin{array}{ccccc}
	A        & \rightarrow & B & + & C  \\
	b \bar b &             & b \bar q &  & q \bar b
	\end{array} \mbox{ },
\end{equation}
where $q$ is a light quark ($u$, $d$ or $s$).
The decay widths were computed as \cite{charmonium,charmonium02,bottomonium,Ferretti:2013vua,3P0,Ackleh:1996yt}: 
\begin{subequations}
\label{eqn:decay-3P0}
\begin{equation}
	\Gamma_{A \rightarrow BC} = \Phi_{A \rightarrow BC}(q_0) \sum_{\ell, J} 
	\left| \left\langle BC \vec q_0  \, \ell J \right| T^\dag \left| A \right\rangle \right|^2 \mbox{ }.
\end{equation}
Here, $T^\dag$ is the pair-creation operator of the $^3P_0$ model, where we substituted the constant pair-creation strength, $\gamma_0$, with an effective one, $\gamma_0^{\mbox{eff}}$, to suppress unphysical heavy quark pair-creation \cite{bottomonium,Ferretti:2013vua,charmonium,charmonium02,Kalashnikova:2005ui}, and introduced a Gaussian quark form factor, because the pair of created quarks has an effective size \cite{bottomonium,Ferretti:2013vua,charmonium,charmonium02,Geiger:1989yc,Geiger-Isgur,SilvestreBrac:1991pw}. $\Phi_{A \rightarrow BC}(q_0)$ is the standard relativistic phase space factor \cite{charmonium,charmonium02,bottomonium,Ferretti:2013vua,3P0,Ackleh:1996yt}, 
\begin{equation}
	\label{eqn:relPSF}
	\Phi_{A \rightarrow BC} = 2 \pi q_0 \frac{E_b(q_0) E_c(q_0)}{M_a}  \mbox{ },
\end{equation}
\end{subequations}
which depends on the relative momentum $q_0$ between $B$ and $C$ and on the energies of the two intermediate state mesons, $E_b = \sqrt{M_b^2 + q_0^2}$ and $E_c = \sqrt{M_c^2 + q_0^2}$. 
The values of the masses $M_a$, $M_b$ and $M_c$, used in the calculation, were taken from the PDG \cite{Nakamura:2010zzi} and Ref. \cite{Mizuk:2012pb}, while, in the case of still unobserved states, we used the predictions from our re-fit of Godfrey and Isgur's model mass formula of Ref. \cite{Ferretti:2013vua}.

%%%%%%%%%%%%%%%%%%%%%%%%%%%%%%%%%%%%%%%%%%%%%%%%%%%%%%%%%%%%%%%%%%%%%
\begin{table*}
\centering
\begin{tabular}{lllll} 
\hline 
Transition & $E_{\gamma}(\mbox{QM})$ (MeV) & $\Gamma_{b \bar b}(\mbox{QM})$ (KeV) & $E_{\gamma}(\mbox{UQM})$ (MeV) & 
            $\Gamma_{b \bar b}(\mbox{UQM})$ (KeV) \\  
\hline       
$\chi_{b0}(3^3P_0) \rightarrow \Upsilon(1^3S_1) \gamma$     & 983                &   0.6         & 984                               & 0.2 \\ 
$\chi_{b0}(3^3P_0) \rightarrow \Upsilon(2^3S_1) \gamma$     & 460                &   1.2         & 461                               & 0.7 \\
$\chi_{b0}(3^3P_0) \rightarrow \Upsilon(3^3S_1) \gamma$     & 138                &    6.1        & 139                               & 6.4 \\
$\chi_{b0}(3^3P_0) \rightarrow \Upsilon(1^3D_1) \gamma$     & 344                &   0.2        & 376                               & 0.2 \\ 
$\chi_{b0}(3^3P_0) \rightarrow \Upsilon(2^3D_1) \gamma$     & 69                 &   0.9         & --                                  & --  \\                                                                                         
$\chi_{b1}(3^3P_1) \rightarrow \Upsilon(1^3S_1) \gamma$     & 971                 &   2.1         & 1061                            & 1.7 \\ 
$\chi_{b1}(3^3P_1) \rightarrow \Upsilon(2^3S_1) \gamma$     & 477                 &   2.5       & 542                              & 2.7 \\
$\chi_{b1}(3^3P_1) \rightarrow \Upsilon(3^3S_1) \gamma$     & 155                 &    7.4         &  223                             & 22.8 \\
$\chi_{b1}(3^3P_1) \rightarrow \Upsilon(1^3D_1) \gamma$    & 361                 &     0          & 458                              & 0 \\
$\chi_{b1}(3^3P_1) \rightarrow \Upsilon(2^3D_1) \gamma$    & 86                   &     0.4       & --                                & -- \\
$\chi_{b1}(3^3P_1) \rightarrow \Upsilon_2(1^3D_2) \gamma$ & 341                &      0         & 408                              & 0.1  \\
$\chi_{b1}(3^3P_1) \rightarrow \Upsilon_2(2^3D_2) \gamma$ & 79                  &     1.0       & --                                 & -- \\  
$\chi_{b2}(3^3P_2) \rightarrow \Upsilon(1^3S_1) \gamma$     & 1010              &   3.9         & 1059                             & 3.6 \\ 
$\chi_{b2}(3^3P_2) \rightarrow \Upsilon(2^3S_1) \gamma$     & 489                &   3.8         & 540                               & 4.0 \\
$\chi_{b2}(3^3P_2) \rightarrow \Upsilon(3^3S_1) \gamma$     & 168                &   8.2         &  221                              & 19.6 \\
$\chi_{b2}(3^3P_2) \rightarrow \Upsilon(1^3D_1) \gamma$     & 373               &   0            & 456                               & 0 \\ 
$\chi_{b2}(3^3P_2) \rightarrow \Upsilon(2^3D_1) \gamma$     & 99                 &   0            & --                                 & -- \\ 
$\chi_{b2}(3^3P_2) \rightarrow \Upsilon_2(1^3D_2) \gamma$ & 354                &   0           & 406                               & 0 \\ 
$\chi_{b2}(3^3P_2) \rightarrow \Upsilon_2(2^3D_2) \gamma$ & 92                 &   0.3         & --                                  & --\\ 
$\chi_{b2}(3^3P_2) \rightarrow \Upsilon_3(1^3D_3) \gamma$ & 357               &   0            & 441                                & 0 \\ 
$\chi_{b2}(3^3P_2) \rightarrow \Upsilon_3(2^3D_3) \gamma$ & 86                 &   1.4         & --                                  & --\\ 
\hline
\end{tabular}
\caption{$E1$ radiative transitions of $\chi_b(3P)$ states, calculated with Eq. (\ref{eqn:radiative}) \cite{charmonium02}. In the calculation, we used our QM and UQM predictions for the masses of the $\chi_b(3P)$'s \cite{Ferretti:2013vua} and the wave functions of the relativized QM \cite{Godfrey:1985xj}. Table from Ref. \cite{charmonium02}; APS copyright.}
\label{tab:e.m.-decays-chiB}  
\end{table*}
%%%%%%%%%%%%%%%%%%%%%%%%%%%%%%%%%%%%%%%%%%%%%%%%%%%%%%%%%%%%%%%%%%%%%

\section{QM calculation of $\chi_b(3P)$'s $E1$ radiative transitions}
In this section, we show our QM results for the radiative transitions of $\chi_b(3P)$ states from Ref. \cite{charmonium02}. The $E1$ radiative transitions were calculated according to \cite{Eichten:1978tg}:
\begin{subequations}
\label{eqn:radiative}
\begin{equation}
	\Gamma_{E1} = \frac{4}{3} C_{fi} \delta_{SS'} e_b^2 \alpha \left| \left\langle \psi_f \right| r \left| \psi_i \right\rangle \right|^2 E_\gamma^3 
	\frac{E_f^{(b \bar b)}}{M_i^{(b \bar b)}}  \mbox{ }.
\end{equation}
Here, $e_b = -\frac{1}{3}$ is the charge of the $b$-quark (in units of $e$), $\alpha$ is the fine structure constant, $E_\gamma$ is the final photon energy, $E_f^{(b \bar b)}$ is the total energy of the final $b \bar b$ state, $M_i^{(b \bar b)}$ is the mass of the initial $b \bar b$ state, $\left\langle \psi_f \right| r \left| \psi_i \right\rangle$ is a radial matrix element, which involves the initial and final radial wave functions, and the angular matrix element $C_{fi}$ is given by
\begin{equation}
	C_{fi} = \mbox{max} (L,L') (2J'+1) \left\{ \begin{array}{rcl} L' & J' & S \\ J & L & 1\end{array} \right\}^2 \mbox{ }.
\end{equation}
\end{subequations}
We calculated the matrix elements of Eqs. (\ref{eqn:radiative}) assuming, for the initial and final states, the wave functions of Godfrey and Isgur's relativized QM \cite{Godfrey:1985xj}. 
We gave results considering, for the masses of the decaying $\chi_b(3P)$ resonances, predictions from the UQM and our re-fit of the relativized QM mass formula of Ref. \cite{Ferretti:2013vua}. 
This was done because the phase space factor of the decay strongly depends on the masses of the initial and final state mesons and our UQM and relativized QM predictions for the masses of $\chi_b(3P)$ states are quite different.

Finally, our results are reported in Table \ref{tab:e.m.-decays-chiB}.

\section{Conclusion}
In this contribution, we discussed the results of an UQM calculation of the bottomonium spectrum with self-energy corrections \cite{Ferretti:2013vua}. 
In the UQM \cite{bottomonium,Ferretti:2013vua,Bijker:2009up,Santopinto:2010zza,Bijker:2012zza,charmonium,charmonium02}, the effects of $q \bar q$ sea pairs are introduced explicitly into the quark model (QM) via a QCD-inspired $^{3}P_0$ pair-creation mechanism \cite{3P0}. 
The $q \bar q$ pairs have vacuum quantum numbers and their effect on hadron observables can be added as a perturbation.
Our UQM results are similar to those of the relativized QM \cite{Godfrey:1985xj} for the majority of states we considered, but differ substantially in the case $\chi_b(3P)$ mesons.
This happens because $\chi_b(3P)$ states are very close to the first open-bottom decay thresholds, thus meson-meson continuum effects can play an important role. 
A comparison between UQM and relativized QM results for the mass barycenter of $\chi_b(3P)$ states was done but, up to now, the present experimental data do not permit to conclude which of the two interpretations is preferable. 
Finally, we discussed our results for the open-bottom strong decay widths of $b \bar b$ states within the $^3P_0$ pair-creation model \cite{Ferretti:2013vua} and for the radiative transitions of $\chi_b(3P)$ states in the QM formalism \cite{charmonium02}. 

In Refs. \cite{Guo:2007sm}, the QM was extended the include the quark-antiquark-gluon higher Fock components to study the spectroscopy of hybrid mesons. In Ref. \cite{Santopinto:2006my} can be found a classification of possible tetraquark states.

\end{document}